\documentclass[a4paper]{article}
\usepackage[T1]{fontenc}
\usepackage[english]{babel}
\usepackage{float}
\usepackage{bm}
\usepackage{amsmath}
\usepackage{amsthm}
\usepackage{graphicx}
\usepackage{amsfonts}
\usepackage[]{color}
\usepackage{fancybox}
\newtheorem{proposition}{Proposition}[section]

\makeatletter

\DeclareTextSymbolDefault{\textquotedbl}{T1}

\numberwithin{equation}{section}
\numberwithin{figure}{section}

\usepackage[all]{xy}
\usepackage{graphicx}

\makeatother

\begin{document}
\begin{flushleft} {\bf OUJ-FTC-20}\\ {\bf OCHA-PP-383}\\ \end{flushleft}

\begin{center} 

{\LARGE{}
Reduction of Complex Dynamics in Far-from-equilibrium Systems: Nambu Non-equilibrium Thermodynamics
}{\LARGE\par}

\vspace{32pt} 

So Katagiri\textsuperscript{{*}}\footnote{So.Katagiri@gmail.com},
Yoshiki Matsuoka$^{*}$ \footnote{machia1805@gmail.com}, and Akio Sugamoto$^{\dagger}$ \footnote{sugamot.akio@ocha.ac.jp}
\begin{center}
\textit{$^{*}$Nature and Environment, Faculty of Liberal Arts, The
Open University of Japan, Chiba 261-8586, Japan}
\par\end{center}

\begin{center}
\textit{$^{\dagger}$Department of Physics, Graduate School of Humanities
and Sciences, Ochanomizu University, 2-1-1 Otsuka, Bunkyo-ku, Tokyo
112-8610, Japan }\\
\par\end{center}

\vspace{10pt}

\end{center}

\begin{abstract}
Far-from-equilibrium thermodynamic systems dominated by strong nonlinearity are reformulated within a dynamical framework based on the Nambu bracket formalism.
It is demonstrated that general complex nonlinear non-equilibrium systems can be locally reduced to a simple form of Nambu Non-equilibrium Thermodynamics (NNET).
Furthermore, mathematical and dynamical obstacles encountered in extending this reduction globally are discussed, and a generalized formulation that incorporates nonlinear effects through mixed higher-order tensors is proposed.
\end{abstract}

\section{Introduction }
In non-equilibrium thermodynamics far from equilibrium, various nonlinear effects play a crucial role. As nonlinearity becomes stronger, even small fluctuations can exceed a certain threshold and produce qualitative changes in the system’s behavior. Phenomena such as periodic motion and pattern formation emerge precisely as a result of such nonlinear effects \cite{Prigogine}. It has long been difficult, however, to discuss such features within the framework of conventional non-equilibrium thermodynamics.
\\
In our previous study \cite{paper 0}\cite{paper I}, we proposed to describe non-equilibrium thermodynamics as a dynamics combining oscillatory (Hamiltonian-like) and dissipative parts, by introducing a set of $N$ thermodynamic variables $x^{i}(t)$ ($i = 1, \dots, N$),
$N - 1$ Hamiltonians $H_{1}, \dots, H_{N-1}$, and an entropy $S$:

 \begin{equation}
\dot{x}^i(t)=\{x^i,H_{1}(\bm{x}),\dots,H_{N-1}(\bm{x})\}_{NB}+\delta^{ij}\partial_{j}S(\bm{x}). \label{autonomous system}
\end{equation}
We called this framework \emph{Nambu Non-equilibrium Thermodynamics (NNET)}.
Here, the Jacobian is called ``Nambu bracket (NB)''\cite{Nambu} in the $N$ dimensional phase space $\{x^1, \cdots, x^N\}$, which is an extension of the Poisson bracket in the usual two dimensional phase space $\{x, p\}$: 
\begin{equation}
  \{A_{1}, \cdots, A_{N}\}_{NB}\equiv\epsilon^{i_{1} \cdots i_{N}}\frac{\partial A_{1}}{\partial x^{i_{1}}} \cdots \frac{\partial A_{N}}{\partial x^{i_{N}}}.
\end{equation}

Its special case for a multiple set of three ($N=3$) dimensional phase space, was found by A. Clebsch\cite{Clebsch}.  Clebsch took the vorticity field as a variable, while Nambu did the velocity field.  Both theories are \emph{dually related}, giving the same equation of motions, under the replacement of vorticity by velocity, and Clebsch functions by Nambu Hamiltonians. See \cite{Nambu2-3}.

The right-hand side of Eq. (\ref{autonomous system}) is the velocity field of a $N$ dimensional fluid, which consists of two terms, $v^{i}(\bm{x})=(\bm{v}^{(1)}(\bm{x})+\bm{v}^{(2)}(\bm{x}))^{i}$; the first term given by multiple Hamiltonians describes an \emph{incompressible} fluid, while the second one written by entropy describes a \emph{compressible} fluid.  The hydrodynamics in terms of Nambu dynamics is developed by Nambu in his last work \cite{Nambu2-3}\cite{Nambu2-1}\cite{Nambu2-2}. 

Given a NNET, it can be a $N$-dimensional autonomous system, having incompressible $\bm{v}^{(1)}(\bm{x})$ and compressive $\bm{v}^{(2)}(\bm{x})$ flows.  Therefore, the proposition that a NNET can be an autonomous system with two kinds of flows is correct.  However, the inverse proposition—that a NNET can be derived from a given autonomous system—is not evident. In this paper we present formal proof, based on the theorems of Helmholtz, Hodge, and Darboux, while leaving the rigorous examination of its necessary conditions to future work.

The formal proof ignores the various obstacles which may appear to destroy the theorem.  About these obstacles we will give a discussion in the last section, in comparison with the global non-existence of the first integral.  In the paper, we illustrate the reduction with representative examples on the reduction of non-linear non-equilibrium systems to a NNET.
\\

In the next section, we study the behavior of systems far from equilibrium, starting from a linear response theory to the non-linear response theory.  In Section three the more complex non-linear non-equilibrium system is examined.  The existence proposition of NNET is proved in Section four, which enables the reduction of a given complex non-linear non-equilibrium system to a simple NNET\footnote{Various examples of the reduction found in \cite{paper I} and \cite{paper III} .}.    
Section five is the summary of this paper.  The last section seven is devoted to the discussions on these unsolved issues.

In Appendix A, we discuss the physical interpretation of Darboux’s theorem. Appendix B illustrates explanations of several examples of the reduction. Appendix C presents detailed discussions of the Helmholtz decomposition and Darboux’s theorem in the case of a triangular chemical reaction. Appendix D analyzes the time evolution of entropy production and the linear stability of steady and cyclic states in Nambu Non-equilibrium Thermodynamics.

\section{Nambu non-equilibrium thermodynamics (NNET) far from equilibrium}

In understanding the behavior of systems far from equilibrium, entropy plays a crucial role.

\subsection{Change of entropy}
The original equation of motion in (\ref{autonomous system}) of the NNET consists of two flows: one incompressible and the other compressible flows.
In the same manner, change of entropy $dS$ also consists of two terms: one related to the \emph{inflow and outflow} of entropy which follows Nambu dynamics, and the other giving the \emph{generation} of entropy inside by dissipation.  We denote the first term by $d^{(H)}S$, and the second by $d^{(S)}S$ following \cite{paper I}:
\begin{equation}
dS=d^{(H)}S+d^{(S)}S,
\end{equation}
which is explicitly given by
\begin{align}
d^{(H)}O = dt \{H_1, \cdots, H_{N-1}, O(\bm{x})\}_{NB}, ~\text{and}~~d^{(S)}O = dt \nabla  O(\bm{x}) \cdot \nabla S.
\end{align}
The first term $d^{(H)}S$ can be applicable to various processes of heat flow, mass transfer and of radiative transfer. The second term $d^{(S)}S$ arises in the irreversible processes, having dissipation and relaxation, which induces the increase of the entropy of the system, since
\begin{align}
\frac{d^{(S)}S}{dt}= (\nabla  S(\bm{x}))^2 \ge 0.
\end{align}
This is the second law of thermodynamics.

Next, we consider a cyclic process (cyclic engine) for which the sum $dS$ becomes a total derivative  \footnote{For the cyclic engine, the state returns to the original one after the motion of one cycle.  The total derivative condition for $dS$ holds, if the generated entropy in a cycle flows out to the outside, but if not, the condition is violated.}, and hence the following integral around a cycle $C$ yields 
\begin{equation}
0=\oint_C dS=\oint_C \left(d^{(H)}S+d^{(S)}S\right),
\end{equation}
so 
\begin{equation}
-\oint d^{(S)}S=\oint_C d^{(H)}S=\oint_C \frac{dQ}{T}
\end{equation}
where $dQ$ is the heat flowing in from outside to the system, and $T$ is the temperature of the system.

\subsection{Generalization of formula in Carnot cycle}
The cycle describes a cyclic heat engine (Carnot cycle and its extension), for which the temperatures are not necessarily two kinds, high $(T_1)_0$ and low $(T_2)_0$.  So, for a general engine, we separate the infinitesimal heat flows $dQ_i~(i=1, 2, \cdots)$ into two sets, the positive (coming-in) set $P$ and the negative (going-out) set $N$, and define the \emph{averaged high temperature} $1/T_1$ and the \emph{averaged low temperature} $1/T_2$ as follows:
\begin{align}
\frac{Q_1}{T_1} = \sum_{i \in P} \frac{|dQ_i|}{T_i}, ~\text{and}~\frac{Q_2}{T_2} = \sum_{i \in N}   \frac{|dQ_i|}{T_i},
\end{align}
where $T_i$ is the temperature when an infinitesimal heat $dQ_i$ flows in, and the total heats flowing in and out are $Q_1=\sum_{i \in P}  dQ_i$ and $Q_2= -\sum_{i \in N}  dQ_i$, respectively.

Then, we have for any cyclic engine a simple relation
\begin{align}
\oint \frac{dQ}{T} = \frac{Q_1}{T_1} - \frac{Q_2}{T_2},
\end{align}
a little refined formula from Carnot cycle.  Using the energy conservation $Q_1 - Q_2=W$, with $W$ the work done and the efficiency $\eta=\frac{W}{Q_1}$, we obtain
\begin{align}
\oint \frac{dQ}{T}  =  \frac{Q_1}{T_2}\left(\eta- \frac{T_1-T_2}{T_1} \right). \label{efficiency}
\end{align}
Therefore, we have  
\begin{align}
\left[ \oint d^{(S)} S \ge 0 \right]  \Leftrightarrow \left[ \eta \le \frac{T_1-T_2}{T_1} \right].
\end{align}
For the special case of $\eta=W=0$, we have $[T_1 \ge T_2 ]$.  Eq.(\ref{efficiency}) implies the entropy production $\oint d^{(S)} S \ge 0$, in terms of \emph{averaged inverse temperatures} $1/T_1$ and $1/T_2$.  For $W=0$, we can say the entropy production is consistent with the \emph{nature of heat};  the heat flows from the higher temperature parts \emph{in average} to the lower temperature parts \emph{in average}.

\color{black}
\begin{table}[t]
\centering
\caption{Summary of symbols and brackets used in Sections 2--3.
The table also illustrates the flow from microscopic variables
to generalized Hamiltonian--entropy structures.}
\label{tab:notation-summary}
\renewcommand{\arraystretch}{1.3}
\begin{tabular}{|p{5.2cm}|p{9.0cm}|}
\hline
\textbf{Name} & \textbf{Definition} \\
\hline\hline
Nambu bracket &
$\{A_{1},A_{2},\dots,A_{n}\}_{\mathrm{NB}}
=\epsilon^{i_{1}\dots i_{n}}
\frac{\partial A_{1}}{\partial x^{i_{1}}}
\frac{\partial A_{2}}{\partial x^{i_{2}}}
\dots
\frac{\partial A_{n}}{\partial x^{i_{n}}}$ \\ \hline

Anti-symmetric bracket &
$\{A_{1},A_{2}\}_{B}
= B^{ij}
\frac{\partial A_{1}}{\partial x^{i}}
\frac{\partial A_{2}}{\partial x^{j}}$ \\ \hline

Symmetric bracket &
$(A_{1},A_{2})_{L}
= L^{ij}
\frac{\partial A_{1}}{\partial x^{i}}
\frac{\partial A_{2}}{\partial x^{j}}$ \\ \hline

Completely symmetric bracket &
$(A_{1},A_{2},\dots,A_{n})_{L_{n}}
= L^{i_{1}\dots i_{n}}
\frac{\partial A_{1}}{\partial x^{i_{1}}}
\frac{\partial A_{2}}{\partial x^{i_{2}}}
\dots
\frac{\partial A_{n}}{\partial x^{i_{n}}}$ \\ \hline

Anti-symmetric tensor &
$B^{ij} = - B^{ji}$ \\ \hline

Transport coefficient
(second-order symmetric tensor) &
$L^{ij} = L^{ji}$ \\ \hline

Completely symmetric tensor &
$L^{i_{1}\dots i_{n}}
= L^{i_{\sigma(1)}\dots i_{\sigma(n)}},
\quad \sigma \in S_{n}$ \\ \hline

Thermodynamic variables
($x^{i} \in \mathcal{M}$) &
$x^{1},\dots,x^{N}$ \\ \hline

Generalized metric on $\mathcal{M}$ &
$g^{i,i_{1}i_{2}\dots i_{k}}(x)$ \\ \hline

Hamiltonian / entropy variables
($X^{a} \in \tilde{\mathcal{M}}$) &
$X^{1}(x),\dots,X^{G}(x)$ \\ \hline

Metric on $\tilde{\mathcal{M}}$ &
$G_{a_{1}\dots a_{k}}(X)
= G_{a_{\sigma(1)}\dots a_{\sigma(k)}}(X),
\quad \sigma \in S_{k}$ \\ \hline
\end{tabular}
\end{table}\color{black}

\subsection{Entropy flow}

When the system is reversible and isolated, we have 
\begin{equation}
\frac{d^{(H)}S}{dt}=\frac{\partial S}{\partial x^{i}}\frac{d^{(H)}x^{i}}{dt}=\sum_i \frac{\partial S}{\partial x^{i}} \{x^i, H_1, \cdots, H_{N-1}\}_{NB} =  0.
\end{equation}

To satisfy this condition, a sufficient condition is to introduce
the antisymmetric tensor $B_{ij}(x)$, and to write the current,  $d^{(H)}x^{i}/dt$ in the following form 
\begin{equation}
\frac{d^{(H)}x^{i}}{dt}=B^{ij}(x)\frac{\partial S}{\partial x^{j}}. \label{reversible part of current}
\end{equation}
Written in this way, the time evolution of the reversible part of entropy is shown to vanish, since

\begin{equation}
  \frac{d^{(H)}S}{dt}=B^{ij}(x)\frac{\partial S}{\partial x^{i}}\frac{\partial S}{\partial x^{j}}\equiv\{S,S\}_{B}=0. 
\end{equation}
This guarantees vanishing of the time evolution of the reversible part of entropy, which is consistent with the second law of thermodynamics.  In other words, the entropy of a closed system undergoing reversible
changes remains \emph{constant} over time, while irreversible changes lead to an increase in entropy.
Therefore, we will choose one of the Hamiltonians equal to the entropy,
\begin{align}
H_{N-1}=S,
\end{align}
and define the bracket $\{S,O\}_{B}$ as
\begin{align}
&\{S,O\}_{B} \equiv \{O, H_{1},\dots, H_{N-2},S\}_{NB} 
= \frac{\partial S}{\partial x^i} B^{ij}(x) \frac{\partial O}{\partial x^j}, \\
&\text{with}~~B^{ij}(x) = \epsilon^{i_{1}\dots, i_{N-2}ij}\frac{\partial H_{1}}{\partial x^{i_{1}}}\dots\frac{\partial H_{N-2}}{\partial x^{i_{N-2}}}. \label{eq:qualconditionH}
\end{align}

\subsection{Linear response: non-equilibrium close to equilibrium}
Linear response is a prototype of the irreversible process, in which the time evolution of the thermodynamic variables satisfies the following constitutive equation,
\begin{equation}
\frac{d^{(S)}x^{i}}{dt}=L^{ij}\frac{\partial S}{\partial x^{j}}\equiv L^{ij}A_{j},
\end{equation}
where $L^{ij}$ is transport coefficient, a second-order symmetric tensor, and $A_{i}$ is the affinity force.

To realize this linear response, the NNET is modified by replacing $\delta^{ij}$ to the kinetic constant $L^{ij}$, or the metric in the thermodynamic space $\mathcal{M}$  \footnote{The kinetic constant $L^{ij}(x)$ and its inverse $R_{ij}(x)$ are naturally considered as the metric tensor $g^{ij}(x)$ and $g_{ij}(x)$ depending on $x$, in the space $\mathcal{M}$, spanned by the thermodynamic variables $(x^1, \cdots, x^{N})$.\cite{gravity analog} }
\begin{align}
\dot{x}^i(t)=\{x^i ,H_{1}(\bm{x}),\dots,H_{N-1}(\bm{x})\}_{NB}+ L^{ij}\partial_{j}S(\bm{x}). 
\end{align}

The corresponding expression in the reversible process is Eq.(\ref{reversible part of current}), 
\begin{equation}
\frac{d^{(H)}x^{i}}{dt}=B^{ij}(x)\frac{\partial S}{\partial x^{j}} = B^{ij}A_{j}.
\end{equation}

Thus, the symmetric $L^{ij}=L^{ji}$ and the anti-symmetric $B^{ij}=-B^{ji}$ specify the irreversible process with entropy production and the reversible process by the Nambu dynamics, respectively.
As in the reversible case, in the irreversible
case, we will use another bracket notation as well,
\begin{align}
&\frac{d^{(S)}O}{dt}=L^{ij}\frac{\partial S}{\partial x^{i}}\frac{\partial O}{\partial x^{j}}\equiv(S,O)_{L_{2}},~\text{as well as} \\
&\frac{d^{(H)}O}{dt}=B^{ij}\frac{\partial S}{\partial x^{i}}\frac{\partial O}{\partial x^{j}}\equiv \{S,O\}_{B},
\end{align}
where $(x^{i},x^{j})_{L_{2}}=L^{ij}$.

In summary, the thermodynamics of non-equilibrium systems close to
equilibrium can be described in terms of NNET as follows:
\begin{equation}\label{eq:close to equilibrium}
\frac{dO}{dt}=\{O, H_{1},\dots,H_{N-2},S\}_{NB}+(S,O)_{L_{2}}, ~\text{and}~
\frac{dS}{dt}=(S,S)_{L_{2}}.
\end{equation}

\subsection{Non-linear response: Non-equilibrium far from equilibrium}

The important point is, as will be shown below that the NNET is easily extended to systems far from equilibrium.

In such systems, the equations of motion or the constitutive equations, are \emph{not linear} in the affinity force, but in general non-linear with higher-order transport coefficients.  That is, they are described by the following general form,
\begin{equation}
\frac{dO}{dt}=\{O, H_{1},\dots,H_{N-2},S\}_{NB}+\sum_{i=0}^{\infty}(\stackrel{i+1}{\overbrace{S,\dots,S}},O)_{L_{i+2}}.
\end{equation}
The notation introduced in the above represents the non-linear response, where $(i+1)$ product of affinity forces $A_a(x)$ with a kinetic constant $L^{a_1 \cdots a_{j+1}}$ gives the ``thermodynamic current''. 
\begin{align}
& (\stackrel{i+1}{\overbrace{S,\dots,S}},O)_{L_{i+2}}  \equiv \frac{d^{(S)}x^{a_{i+2}}}{dt} \frac{\partial O}{\partial x^{a_{i+2}}}, ~\text{with} ~A_a(x)=\frac{\partial S}{\partial x^a},  \\
&\text{and}~\frac{d^{(S)}x^{a_{i+2}}}{dt}= \sum_{a_1,a_2, \cdots, a_{i+1} }L^{a_1a_2 \cdots a_{i+1}a_{i+2} } A_{a_1}(x) A_{a_2}(x) \cdots A_{a_{i+1}}(x),
\end{align}
where $L^{a_1a_2 \cdots a_{i+1} a_{i+2} }$ are $(i+2)$-order complete symmetric tensors.

Here we will relax the restriction of isolated system and include the ``open systems'' which do not necessarily go to the equilibrium. Therefore, we have to write the reversible
part more carefully, by using a general $H_{N-1}$ instead of $S$, as follows:
\begin{equation}
\frac{dO}{dt}=\{O, H_{1},\dots,H_{N-2},H_{N-1}\}_{NB}+\dots.
\end{equation}
The important point in this choice is that $d^{(H)} x^i/dt$ takes the usual form of the Nambu dynamics,
\begin{eqnarray}
\frac{d^{(H)}x^i}{dt}= B^{ij}(x) \frac{\partial H_{N-1}}{\partial x^j} = \{x^i, H_1, \cdots, H_{N-1} \}_{NB},
\end{eqnarray}
so that $d^{(H)}S/dt \ne 0$. This situation is favorable in open systems, where entropy fluxes are unbalanced and sustained by the environment. \footnote{
This is the crucial difference between systems close to equilibrium and those far from equilibrium. In systems close to equilibrium, as discussed above in (\ref{eq:close to equilibrium}), one of the Hamiltonians can be chosen to $S$ and $d^{(H)}S/dt=0$.
\color{black}
}

\section{Reduction of a complex system to a simple NNET}
Next, we discuss a possibility that such a non-linear system does not move toward the usual equilibrium state, but rather, approaches a new ``cyclic system'' (denoted by $\Omega_{\mathrm{cycle}}$) in the NNET.  We will explain this in the following.  

In the NNET, the general non-linear response can be described by
\begin{eqnarray}
\dot{x}^{j}= \{ x^j, H_1, H_2, \cdots, H_{N-1} \}_{NB} +  
\sum_{i=0}^{\infty} (\stackrel{i+1}{\overbrace{S,\dots,S}}, x^{j})_{L_{i+2}}. \label{3.1}
\end{eqnarray}
The second term in the right-hand side of (\ref{3.1}) has a complicated expression with the multiple product $\prod_{n=1}^{i+2}\partial S/ \partial x^{a_n}$ of vectors with a general transport coefficient $L^{a_1 a_2 \cdots a_{i+2}}$, but the result of the product is a single vector field.  

As was stated in Introduction, we consider the inverse proposition holds; for a given autonomous system with incompressible and compressible flows, we can find a Nambu Non-equilibrium thermodynamics (NNET).
Namely, we can identify $\{H'_1, \dots, H'_{N-1} \}$ and a new entropy $S^C$, and obtain the following expression:
\begin{eqnarray}
\sum_{i=0}^{\infty} (\stackrel{i+1}{\overbrace{S,\dots,S}}, x^{j})_{L_{i+2}}= \{x^{j}, H'_1, H'_2, \cdots, H'_{N-1} \}_{NB} + \partial_j S^C.
\end{eqnarray}
Accordingly,  
\begin{eqnarray}
\dot{x}^{j}= \{ x^{j}, H_1, H_2, \cdots, H_{N-1}\}_{NB} +  \{ x^{j}, H'_1, H'_2, \cdots, H'_{N-1} \}_{NB} + \partial_j S^C. \label{3.3}
\end{eqnarray}
The first and the second terms of (\ref{3.3}) satisfies the incompressibility condition, so that the sum of them also satisfies the condition.  Thus, we can find the third set of Hamiltonians $\{ H^C_{1}, H^C_{2}, \cdots, H^C_{N-1} \}$, which yields the following general expression, even for the system far from equilibrium,
\begin{eqnarray}
\dot{x}^{j}= \{ x^{j}, H^C_1, H^C_2, \cdots, H^C_{N-1} \}_{NB} + \partial_j S^C.
\label{xtime}
\end{eqnarray}
For a general observable $O$, we have
\begin{eqnarray}
\dot{O}= \{ O, H^C_1, H^C_2, \cdots, H^C_{N-1} \}_{NB} + (\nabla S^C \cdot \nabla O).
\label{otime}
\end{eqnarray}
This is the \emph{reduction} of a given complex system to a simple NNET.  It is as if the non-linear response drives the system to a simpler one; the attained simple system is a new non-equilibrium thermodynamics, which approaches a new ``equilibrium''.  The new system is described by a a single entropy $S^C$, which is different from the original entropy $S$ ($dS=dQ/T$), not being the stochastic Shannon entropy nor the fluctuating Onsager-Machlup-Hashitsume entropy.  After arriving at the new simple system $\Omega^C$, the following conditions hold
\begin{eqnarray}
\frac{d S^C}{dt} \ge 0, ~\text{and}~\frac{d^2 S^C}{dt^2} < 0.
\end{eqnarray} 
Under this condition, after a certain period, the system is relaxed to the cyclic system $\Omega^C$.  The reason why we guess the cyclic system is that the new system is strongly controlled by a pure Nambu dynamics with \emph{maximum number of conserved quantities} $(H^C_1, H^C_2, \cdots, H^C_{N-1})$.  Since the conservation laws are violated by the existence of entropy $S^C$, the appearance of the cyclic motion cannot be proved rigorously, but it is suggested in many cases.     

\color{black}
\subsection{Physical interpretation of the multiple Hamiltonians and entropy}
\label{subsec:physical-interpretation}

We now comment on the physical meaning of the multiple Hamiltonians \( H_I^{C} \) and the entropy-like potential \( S^{C} \). For simplicity, we restrict ourselves to the case \( N = 3 \).

In our construction, the velocity field is decomposed as
\begin{equation}
\dot{x}^i = V^{(H)i} + \nabla^i S^{C},
\qquad
V^{(H)i} = \{ x^i, H_1^{C}, H_2^{C} \}_{\mathrm{NB}},
\end{equation}
where \( V^{(H)} \) represents the Nambu (solenoidal) component and \( \nabla S^{C} \) the dissipative component.

Assuming the standard Euclidean volume element \( d^N x \), the incompressibility of Nambu dynamics implies
\begin{equation}
\nabla \cdot V^{(H)} = 0.
\end{equation}
As a result, the local phase-space contraction rate is given by
\begin{equation}
\nabla \cdot \dot{x} = \Delta S^{C}.
\end{equation}
Therefore, \( S^{C} \) can be interpreted as a scalar potential whose Laplacian directly characterizes the local compressibility of the dynamics. Importantly, this quantity can be computed numerically from measured velocity-field data.

The functions \( H_I^{C} \) determine the solenoidal component \( V^{(H)} \). Since
\begin{equation}
V^{(H)} \cdot \nabla H_I^{C} = 0,
\end{equation}
the Nambu flow is tangent to the level surfaces \( H_I^{C} = \mathrm{const} \). However, because the full dynamics satisfies
\begin{equation}
\dot{H}_I^{C} = \nabla H_I^{C} \cdot \nabla S^{C},
\end{equation}
the trajectories generally cross these level surfaces, and \( H_I^{C} \) are not conserved quantities in the presence of dissipation. In this sense, \( H_I^{C} \) play the role of local generators of the conservative part of the dynamics.

Once \( S^{C} \) has been reconstructed numerically, the solenoidal component can be extracted from the observed dynamics \( \dot{x}_D^i \) via
\begin{equation}
V^{(H)i} = \dot{x}_D^i - \nabla^i S^{C}.
\end{equation}
The Hamiltonians \( H_I^{C} \) can then be locally inferred by solving
\begin{equation}
V^{(H)} \cdot \nabla H_I^{C} = 0.
\end{equation}

In summary, the procedure
(i) reconstructing \( S^{C} \) from \( \Delta S^{C} = \nabla \cdot \dot{x} \),
(ii) extracting the reversible component \( V^{(H)} = \dot{x} - \nabla S^{C} \),
and
(iii) constructing first integrals \( H_I^{C} \) satisfying \( V^{(H)} \cdot \nabla H_I^{C} = 0 \),
provides a concrete method for passing from a given \( x \)-frame to a local Darboux frame (or \( y \)-frame).

Mathematically, this corresponds to the Helmholtz decomposition
\begin{equation}
V^{(H)i} = \epsilon^{ijk} B_{jk},
\end{equation}
where the closed two-form \( B = B_{ij} \, dx^i \wedge dx^j \) satisfies \( dB = 0 \).
Darboux’s theorem guarantees that, locally, one can choose coordinates
\( H_1(x), H_2(x) \) such that
\begin{equation}
B = dH_1 \wedge dH_2.
\end{equation}
The physical meaning of this construction is discussed in Appendix~A.

The construction of the Darboux frame is generally local; global descriptions may require patching due to topological obstructions or singularities. Furthermore, the Hamiltonians \( H_1 \) and \( H_2 \) are not unique: transformations preserving \( B \) leave a residual freedom. It is physically useful to exploit this freedom to choose \( H_I^{C} \) so that they coincide with manifestly conserved or approximately conserved quantities of the system. In particular, if the condition
\begin{equation}
\nabla S^{C} \cdot \nabla H_I^{C} = 0
\end{equation}
is satisfied, \( H_I^{C} \) become conserved quantities of the full dynamics.

This observation is important: once the dissipative component has been extracted into \( S^{C} \), the remaining dynamics is necessarily Hamiltonian, and its time evolution can be described using physically transparent conserved or approximately conserved quantities.
\color{black}

\subsection{More general non-linear system with mixed tensors}
We call $T_{i_1i_2\cdots}(x)$ a tensor, having indices of derivatives $\partial/\partial x^i$.  So far what we have considered are the totally \emph{anti-symmetric} tensors in the Nambu dynamics and the totally \emph{symmetric} tensors in the non-linear response theory.  But, in general they can be \emph{mixed} tensors.

Thus, we introduce, without specifying Hamiltonian or entropy, $G$ functions $\{ X^1(x), X^2(x), \cdots, X^{G}(x) \}$ for $N$ thermodynamic variables 
$\{x^1, x^2, \cdots, x^N\}$.  Then, the more general fluid dynamics (autonomous system) can be given, in terms of a ``generalized metric'' $g^{i, i_1i_2 \cdots i_k}(x)$ in the thermodynamic space $\mathcal{M}$, ``affinity forces'' (vectors) $A_{i}^a(x)=\partial_{i} X^a$, and another metric $G_{a_1a_2\cdots a_k}(X)$ in the Hamiltonian-Entropy space $\mathcal{\tilde{M}}$, as follows:
\begin{align}
&\dot{x}^i(t) = v^i (x) \nonumber \\
&= \sum_{k=1}^{\infty}\sum_{\{i_1 \cdots i_k\}=1}^N \sum_{\{a_1 \cdots a_k\}=1}^G \; g^{i, i_1i_2 \cdots i_k}(x)  G_{a_1a_2\cdots a_k}(X) A^{a_1}_{i_1}(x)A^{a_2}_{i_2}(x) \cdots A^{a_k}_{i_k}(x).
\end{align}
Here, the generalized metric $g^{i, i_1i_2 \cdots}(x)$, a generalized kinetic constant, is a \emph{mixed tensor} with symmetric and anti-symmetric indices. \footnote{The generalized metric may determine the geometry of the thermodynamic space $\mathcal{M}=\{x^i\}$ (a kind of parameter space). See \cite{paper I} and discussion in Subsection 7.1.  Here, the other metric $G_{a_1a_2\cdots a_k}(X)$ is also introduced in the Hamiltonian-Entropy space $\{X^a\}=\{H_1, H_2, \cdots; S_1, S_2, \cdots \}$, which can be denoted as  $\tilde{\mathcal{M}}$ (a kind of target space).  Here the target space metric is fixed to be a specific one.}

In the same manner as before, we admit the validity of the ``inverse proposition'', which
claims the reduction of the complicated velocity field $v^i(x)$ to a simple NNET. To demonstrate this, we have to find proper Hamiltonians $\{H_1, \cdots, H_{N-1}\}$ and entropy $S$ from the original $\{X^1, \cdots, X^G\}$.  For this purpose, we need the proposition given in the next section.

\color{black}
\begin{figure}[t]
\centering
\Ovalbox{%
\begin{minipage}{1.1\linewidth}
\centering

\begin{equation}
\dot{x}^{i}
= v^{i}(x)
= \sum_{k=1}^{\infty}
\sum_{\{i_{1},\dots,i_{k}\}=1}^{N}
\sum_{\{a_{1},\dots,a_{k}\}=1}^{G}
g^{i,i_{1}\dots i_{k}}(x)\,
G_{a_{1}\dots a_{k}}(X)\,
A_{i_{1}}^{a_{1}}(x)\cdots A_{i_{k}}^{a_{k}}(x)
\end{equation}

\[
\Downarrow
\quad
\text{\rm (Helmholtz/Hodge decomposition on a local patch)}
\quad
\Downarrow
\]

\begin{equation}
\dot{x}^{i}
= B^{ij}\frac{\partial H^{(C)}}{\partial x^{j}}
+ L^{ij}\frac{\partial S^{(C)}}{\partial x^{j}},
\qquad
\text{\rm (divergence-free part + dissipative part)}
\end{equation}

\[
\Downarrow
\quad
\text{\rm (Darboux theorem for the divergence-free part)}
\quad
\Downarrow
\]

\begin{equation}
\dot{x}^{i}
= \{x^{i},H_{1}^{(C)},\dots,H_{N-1}^{(C)}\}_{\mathrm{NB}}
+ L^{ij}\frac{\partial S^{(C)}}{\partial x^{j}}
\end{equation}

\end{minipage}}
\caption{Schematic flowchart of the reduction procedure from a general autonomous dynamical system to the Nambu Non-equilibrium Thermodynamics (NNET) form.}
\label{fig:reduction-flowchart}
\end{figure}
\color{black}

\section{Existence proposition of NNET}
\begin{proposition}
This proposition is mentioned in the above as the ``inverse problem''; it claims that we can find a Nambu Non-equilibrium thermodynamics (NNET) for a given autonomous system, with incompressible $v^{(1)\mu}(x)$ and compressible $v^{(2)\mu}(x)$ flows, acting in the $N$-dimensional space. 
\end{proposition}

\subsection{Local existence proposition of NNET}

This proposition is a modified version of the well-known ``Helmholtz theorem''. In hydrodynamics (or Hodge theorem in mathematics), which states $^\exists \psi^{\mu\nu}(x)$ (2-form field) s.t. 
\begin{equation}
v^{(1)\mu}(x)=\partial_{\nu}\psi^{\mu\nu}(x),
\end{equation}
 and $^\exists \phi(x)$ (0-form field) s.t. 
\begin{equation}
v^{(2)\mu}=\delta^{\mu\nu}\partial_{\nu}\phi(x)
\end{equation}
in the flat $x$-space.\footnote{By the Helmholtz decomposition, an entropy function $S$ can always be introduced locally. However, whether a globally smooth $S$ exists depends on the topology of the phase space and the presence of singularities. This issue remains a subject for future study.}

In general, another velocity field $\bm{v}^{(3)}(x)$, a harmonic one $\nabla^2 \bm{v}^{(3)}(x)=0$, can be added, but here $\bm{v}^{(3)}(x)=0$ holds, on the assumption of no spacial singularities and $\bm{v}^{(3)}(x)$ vanishes at spacial infinity.

Next,  ``Darboux theorem''  \cite{Darboux}  gives a \emph{canonical frame} $\{y^i\}$.  That is, by a proper coordinate transformation, the 2-form field can be reduced to a constant skew symmetric form (\emph{canonical form}),
\begin{equation}
 \psi^{'ij}=\sum_{k=1}^p \left( \delta^i_ {2k-1}\delta^j_{2k}-\delta^i_{2k}\delta^j_{2k-1} \right),
 \end{equation}
 where $^\exists p$ s.t. $2p \le N$.  A physical view of the Darboux theorem is given in Appendix, where $\psi_{ij}$ is considered to be a field strength in electromagnetism.
Then, the corresponding dual $(N-2)$-form in $x$ space can be written by the 2-form in the new frame $y$,
\begin{align}
&\psi^*_{\mu_3 \cdots \mu_N}(x)= \psi'^*_{i_3 \cdots i_N}(y)  \left( \frac{\partial y^{i_3}}{\partial x^{\mu_3}} \cdots \frac{\partial y^{i_N}}{\partial x^{\mu_N}} \right), \\
&\text{with}~\psi'^*_{i_3 \cdots i_N}(y)=\frac{1}{2}  \epsilon'_{ij i_3 \cdots i_N}(y) \psi'^{ij}(y),
\end{align} 
where the epsilon tensor $\epsilon'$ in the new frame may not be a flat space one $\epsilon^{(0)}$, but is a curved space one satisfying
\begin{align}
\epsilon'_{ij i_3 \cdots i_N} = \sqrt{g(y)} \epsilon^{(0)}_{ij i_3 \cdots i_N},
\end{align}
where $g(y)= \text{det} \; g_{ij}(y)$ is a determinant of the metric in the $y$-frame.  This relation can be understood from the volume form.

Substituting the canonical form in $\psi'^{ij}(y)$, we have
\begin{align}
\psi^*_{\mu_3 \cdots \mu_N}(x) =2\sqrt{g(y)}  \epsilon^{(0)}_{2k-1, 2k, i_3 \cdots i_N}  \left( \frac{\partial y^{i_3}}{\partial x^{\mu_3}} \cdots \frac{\partial y^{i_N}}{\partial x^{\mu_N}} \right),   
\end{align}
where $(2k-1, 2k)$ are the remaining indices other than $(i_3, \cdots, i_N$). The epsilon symbol with $(0)$ takes $\pm 1$ according to even and odd permutations of the indices.
Then, we identify Nambu Hamiltonians $(H_1, \cdots, H_{N-1})$ by $y^{i+1} = \ln H_{i}(x)$ and $2\sqrt{g(y)}= \prod_{i=1}^{N-1} H_i(x)$, which yields 
\begin{align}
\psi^*_{\mu_3 \cdots \mu_N}(x)=  \sum_{i_1, \cdots, i_{N-1} \in (1, \cdots, N-1)}
 \epsilon^{(0)}_{i_1 \cdots i_{N-1}}    \left( H_{i_1} \frac{ \partial H_{i_2}}{\partial x^{\mu_3}}  \cdots \frac{\partial H_{i_{N-1}}}{\partial x^{\mu_N}}\right).  
\end{align}
In this way, we arrive at the same expression $\dot{x}^{\mu}(t)$ in Nambu dynamics:
\begin{align}
v^{(1)\mu}(x)=\frac{1}{(N-2)!} \epsilon_{(0)}^{\mu\nu\mu_3 \cdots \mu_N} \partial_{\nu} \psi^*_{\mu_3 \cdots \mu_N} =\{x^{\mu}, H_1, H_2, \cdots, H_{N-1}\}_{NB}. 
\end{align} 
The determination of $H's$ is equivalent to find the Darboux's reduced $y$-frame.  

Associated with the transformation to the new frame, the compressible (dissipation) part of flow is also modified, 
\begin{align}
\bm{v}_{\mu}^{(2)}(x) = \bm{v}_i^{'(2)}(y) \frac{\partial y^i}{\partial x^{\mu}}.
\end{align}
The compressibility condition $\partial_{[\mu, }v_{\nu]}^{(2)}(x)=0$ implies $\partial_{[j, } v^{' (2)}_{i]}(y)=0$, so that we can find a new entropy $S'(y)$, leading to
\begin{align}
\bm{v}^{'(2)}(y)_i =  \frac{\partial S'(y)}{\partial y^i},~\text{and}~~\bm{v}^{(2)\mu}(x) =  \delta^{\mu\nu}\frac{\partial S'(y)}{\partial y^i} \frac{\partial y^i}{\partial x^{\nu}}.
\end{align} 
This simple expression gives the reduction in the new frame of a complex expression in the original frame.  For example, in the non-linear response case of (2.24), we have a complex non-linear expression before the reduction,
\begin{align}
\bm{v}^{(2)\mu}(x)= \sum_{a_1,a_2, \cdots, a_{m+1} }L^{a_1a_2 \cdots a_{m+1}a_{m+2} } \partial_{a_1}S(x) \partial_{a_2}S(x) \cdots \partial_{a_{m+1}}S(x).
\end{align}

We have to remember that whether the existence proposition is valid globally or not, depends on the global existence of the Darboux frame $y$ as well as the global existence of $\psi^{\mu\nu}(x)$ and $\phi(x)$. This issue is discussed in Section 6.2.

\subsection{Global non-existence of first integrals}
It is known that the first integrals do not necessarily exist.  For example, Kowalevskaya (1889) \cite{Kowa}  found a new first integral in a three dimensional top under the gravity in $z$ direction.  Her top is a Nambu dynamics of $N=6$, without dissipation.  So, we expect to have five first integrals or five Nambu Hamiltonians $(H_1, \cdots, H_5)$.  This expectation was proved to hold \cite{Kowa}, only under a specific choice of parameters, the moment of inertia $(A, B, C)$ and the center of mass coordinates $\{x_0, y_0, z_0 \}$, such as
\begin{eqnarray}
A=B=2C,  ~\text{and}~y_0=z_0=0.~\text{(Kowalevskaya)}
\end{eqnarray}
This implies that one of the first integrals (or one Hamiltonian) does not exist in almost all cases.  What is the relation between this fact and the existence proposition of NNET just proved in the above?

The existence proposition of NNET shows the existence of Hamiltonian and entropy as a function of $\bm{x}$.  Of course, for the theorem to hold, a number of conditions have to be imposed.  On the other hand, the non-existence of the first integrals \cite{Kowa} \cite{Yoshida} pays attention to whether the time-development of $\bm{x}=\bm{x}(t)$ gives singularities, at a certain time $t_{\star}$, depending on the initial conditions $\bm{x}(0)$.  The breaking of the Existence proposition of NNET and the global non-existence of the first integrals, seem to be delicately interrelated.

This  issue is briefly examined in Section 6.2 , where the obstacles to destroy the existence proposition of NNET are listed up in relation to the appearance of chaos and fractal.

\color{black}
\paragraph{Physical relevance of local reduction}
\label{par:local-reduction-meaning}

Even when global Hamiltonians \( H_I \) or a global entropy function \( S \) cannot be defined, the existence of locally defined \( H_I \) and \( S \) still has clear physical meaning. Such situations naturally arise in systems with topological obstructions, singularities, or chaotic dynamics.

In these cases, as long as the velocity field is sufficiently smooth on a local patch
\( U \subset \mathbb{R}^N \) that does not include singular points, and the decomposition
\begin{equation}
\dot{x} = V^{(H)} + \nabla S
\end{equation}
is applicable, the locally constructed quantities \( H_I|_{U} \) and \( S|_{U} \) provide a physically meaningful description of the dynamics.

In practice, both physical experiments and numerical analyses are necessarily performed on finite spatial domains and over finite time intervals. As long as the reduced description yields reproducible predictions or characteristic features within such finite domains, it functions as a valid effective description. This point is particularly important for chaotic systems: while global integrals of motion cannot be expected in the long-time limit, local contraction rates
\(
\nabla \cdot \dot{x} = \Delta S
\)
and quasi-invariants associated with the reversible component \( V^{(H)} \) (slow variables) can often be identified on appropriate coarse-graining scales or within finite time windows.

Moreover, patchwise descriptions excluding singular neighborhoods are frequently effective, in close analogy with the use of local coordinate systems in fluid dynamics. When necessary, a more global description can be constructed by patching together the local data
\( \{ H_I^{(\alpha)}, S^{(\alpha)} \} \)
defined on multiple patches \( \{ U_\alpha \} \) via suitable transition relations, leading to an atlas-like representation of the dynamics.
\color{black}

\section{Summary}
In this paper we develop in the framework of Nambu non-equilibrium thermodynamics (NNET), how to study the system of non-equilibrium thermodynamics far from equilibrium.  The NNET is given by the Nambu dynamics with multiple Hamiltonians and the dissipative dynamics with an entropy.

The linear response theory is included easily into the model with a kinetic constant $L^{ij}$ as a metric tensor of the thermodynamic space spanned by $\{x^1, \cdots, x^{N} \}$.  The non-linear response theory is also included by introducing a non-linear response with a generalized kinetic constants $L^{a_1a_2 \cdots}$.  This generalized kinetic constants can be further extended to mixed tensors $g^{i, i_1i_2 \cdots}(x)$, having both symmetric and anti-symmetric indices.

This highly non-linear system far from equilibrium can be reduced to a simple NNET with new Hamiltonians and a new entropy.  This is due to the existence proposition of NNET given in Section 4.  The new system obtained (or reduced) approaches a new equilibrium state, which is described by a simple NNET.  Accordingly, the Nambu non-equilibrium thermodynamics (NNET) is a powerful method to study the Belousov-Zhabotinsky reaction (BZ) giving cyclic chemical reactions, as well as Hindmarsh-Rose (H-R) model to describe the propagation of spiky signal along the nerve, as well as the chaotic system of Lorenz and Chen (see \cite{paper III}).  Thus, it is plausible that the reduced system approaches a cyclic motion.

A relation between the existence proposition of NNET and the global non-existence of the first integral is briefly mentioned. 

\section{Discussions}
The issues which we have not examined well are listed in the followings.

\subsection{(Issue 1) Geometry of the generalized metric of mixed tensor}
The metric transforms under the general coordinate transformation, $x^i \to x^{'i}=^\forall \! f^i(x)$, as
\begin{align}
g'^{j, j_1 \cdots j_k}(x') = \partial_i x'^{j} \partial_{i_1} x'^{j_1} \cdots \partial_{i_k} x'^{j_k} g^{i, i_1 \cdots i_k}(x) .
\end{align}
So far nothing is different from the usual Riemann geometry.  However, the concept of parallel transportation and the definition of the curvature-like tensors differ, depending on the exchange symmetry of the indices.  The metric is classified by the symmetry of the indices, or the representation of the symmetry group of order $k$, $S_k$, which can be done in terms of Young tableau.  So, this is an interesting problem to pursue and seems to be related to the structure formation \cite{Prigogine} in NNET. 

\subsection{(Issue 2) Existence proposition of NNET v.s. Global non-existence of the first integral}
As was mentioned in Subsection 4.1, the global non-existence of the first integral occurs when the trajectory $\bm{x}(t)$ cannot be elongated over the singularity. 

If the singularity appears as an isolated point at $t=t_{s}$, then the solution can be extended over the singularity by deforming the path of time development.  Provided for example $dx \sim a(t-t_s)^{\alpha}dt$, we have
\begin{align}
x- x_s \sim \begin{cases} ~\frac{a}{\alpha +1} (t- t_s)^{\alpha +1} ~\text{for}~ \alpha \ne -1, \\
~a \ln (t-t_s) ~\text{for}~ \alpha = -1,
\end{cases}
\end{align}
so $(x- x_s)$ diverges if $\alpha \le -1$.  If we can find a way to avoid the point singularity, such as $x_s \to x_s + i \epsilon$, or deforming the path along a half circle around $t_s$, then we can extend (elongate) the solution over the singularity,
\begin{align}
(x-x_s)|_{t > t_s} = e^{i \pi (\alpha+1)} (x-x_s)|_{t < t_s}.
\end{align}
As we know from the Landau singularity, the real singular behavior arises when two singularities pinch the path and disturb the deformation of the path. Such problem can be managed, following the $i\epsilon$-prescription in quantum field theory (or equivalently the micro-local calculus in mathematics by Mikio Sato \cite{Sato}).

\subsection{Mathematical obstacles}
The first class of obstacles arises from purely mathematical constraints. These include the non-existence of smooth stream functions, failure to construct a Darboux canonical frame, or topological changes in the thermodynamic space. Such problems are structural, independent of the detailed dynamics.

The obstacles to destroy the proposition can be \\
1) Obstacle to introduce stream function $\psi^{\mu\nu}(x)$ and velocity potential $\phi(x)$. \\
2)  Obstacle to introduce canonical frame of Darboux. \\
The first one 1) is related to the disturbance against the smooth flow, such as the appearance of vortices, turbulence \cite{SLE}, bifurcation and singularities (such as attractors) in the flow.  This is not an easy problem, and is intimately related to fractal and chaos. The appearance of singularities (attractors) in the flow may change the topology of the thermodynamic space $\mathcal{M}$, adding a non-trivial harmonic flow $\bm{v}^{(3)}$.\\
The second one 2) is also a quite difficult problem.  It is essentially the problem in electromagnetism (see Appendix), but the canonical frame is not necessary a flat space, so that the obstacle may arise relating to the topology change of the thermodynamic space $\mathcal{M}$.  This obstacle was studied in the context of symplectic transformation between two manifolds with different topology by Moser and Gromov.\cite{M-G}

\subsection{Dynamical obstacles: chaos and fractals}
Even if the mathematical structure exists locally, dynamical effects may break the reduction. Chaotic behavior, fractal trajectories, and monodromy instabilities in the Poincaré map exemplify this second class of obstacles.

 Thus, dynamical breakdowns of the proposition are closely tied to chaos and fractal structures, and hence, the two problems in (Issue 2) are intimately interrelated.
Treatment of chaos and/or fractal is rather well-known from the monodromy behavior of the Poincar\'{e} map \cite{Poincare}. (See also \cite{Boltzmann})
The Nambu dynamics can be roughly viewed as a periodic motion (cycle), so we can assume the trajectory in the 0-th approximation is a fixed cycle $\bm{x}_C$.  

Then, we can introduce a Poincar\'{e} map \cite{Poincare} for $N$-dimensional phase space, at discrete time steps separated by the period $T$ of the cycle.  In the 0-th approximation, we have
\begin{align}
\bm{x}(nT)_C= \cdots=\bm{x}(T)_C= \bm{x}(0)_C.
\end{align}
To know how the fluctuation (or perturbation) $\bm{\xi}(t)$ deforms the trajectory, we write the Poincar\'{e} map of the fluctuation by a \emph{monodromy} matrix $\hat{M}$:
\begin{align}
\bm{\xi}(T)= \hat{M} \bm{\xi}(0), ~\text{and} ~\bm{\xi}(nT)= (\hat{M})^n \bm{\xi}(0).
\end{align}
Here, the bold faces $\bm{x}_C$ and $\bm{\xi}$ denote  $N$-dimensional vectors.

After $\infty$-times ($n \to \infty$) Poincar\'{e} maps, the fluctuations may pile up and distort the original periodic trajectory $\bm{x}_C$ largely.
The $N \times N$ monodromy matrix $\hat{M}$ is not always diagonalizable.  But, for a meanwhile we forget it and introduce $N$ eigenvalues and eigenvectors as $\lambda_a$ and $\bm{\xi}_a$ ($a=1,\dots , N$).  Then, $n$-times Poincar\'{e} maps gives 
\begin{align}
\bm{\xi}_a(nT) =(\lambda_a)^n \bm{\xi}_a(0),
\end{align}
where the eigenvalues and eigenvectors can be complex number and vectors.

There exist two types of singular behavior which lead to a different motion apart from the cyclic one.  Define the modulus and angle by $\lambda=|\lambda| e^{i\theta}$, then we have (type 1) : the modulus of eigenvalue behaves $|\lambda^n| \to \infty$ for $n \to \infty$, and (type 2): even for the limit cycle with $|\lambda|=1$, $\{ n \theta | n= 1, 2, \cdots, \infty\}$ sweep the whole angle of $2\pi$.

\subsubsection{(type 1): Chaos}
In this case, a tiny fluctuation grows infinitely large, giving a chaos.   The entropy behaves like a Lyapunov index, so it may enhance or depress this chaotic behavior.

\subsubsection{(type 2): Fractal}
This type is well studied by Kolmogorov-Arnold-Moser (KAM) \cite{KAM} or Kowalevskaya \cite{Kowa}. 
Eigen-modes can be classified to two classes; one is the  ``resonant'' eigen-mode, where the angle takes a form $\theta=2\pi (r), (r)$:\emph{rational number}.  The other is the ``non-resonant'' eigen-mode where the angle $\theta= 2\pi(irr)$, $(irr)$:\emph{irrational number}.
The non-resonant case matches (type 2) condition and  the trajectory becomes \emph{fractal} having the fractal dimension greater than 1.  

\subsubsection{ (type 3): Degenerate case in general}
Next, we consider the case of non-diagonal monodromy matrix, having a degeneracy in the eigenvalues.  Even in this case, the matrix $\hat{M}$ can be transformed to a Jordan normal form $\hat{J}$, by finding an invertible matrix $T$:
\begin{align}
T^{-1} \hat{M} T= \hat{J}= \sum_{b=1}^{B} \oplus \hat{J}_b, ~\text{where}~(\hat{J}_b)_{ij}= \lambda_b \delta_{ij} + \delta_{i, i+1}, ~\text{for}~(i, j)=1, \dots ,m_b. \label{Jordan}
\end{align}  
The eigenvalue $\lambda_b$ in the $b$-th Jordan block, is $m_b$-ply degenerate.  Then, the bases of the block $\{\bm{\xi}^b_1, \cdots, \bm{\xi}^b_{m_b}\}$ form the generalized (or weak) eigen-vectors, and $T=[\bm{\xi}^b_1, \cdots, \bm{\xi}^b_{m_b}]$ is given by placing these column vectors side by side. We denote the space spanned by these ``eigen-vectors'' as $\Xi^b$.
From (\ref{Jordan}), we have 
\begin{align}
\hat{M} \xi^b_k = \xi^b_{k-1} + \lambda_b \xi^b_k.
\end{align}
This yields after $n$-times Poincar\'{e} mappings, (we may omit the block index $b$ here), 
\begin{align}
\bm{\xi}_k(nT)=\hat{M}^n \bm{\xi}_k =  \sum_{i \;(\text{for}\; i \ge 0, \; k-n+i \ge 0)}^n \begin{pmatrix} n \\ i \end{pmatrix}  \lambda^i \bm{\xi}_{k-n+i} \in \Xi.
\end{align}
The fluctuation vectors $\{\bm{\xi}_k(nT)\}$ at each Poincar\'{e} map move within the space $\Xi$, where the transverse components against the original vector can be induced.  Therefore, more generally than (type 1: Chaos) and (type 2: Fractal), fractal and/or chaos appear also in this general case, depending on the choice of eigenvalues and generalized eigen-modes for each Jordan block. 

Recently, the global existence of the first integral (conserved quantity) is developed in mathematics, as a theorem by Ziglin (1983) and others.  Refer to \cite{Yoshida}.  These works are done of course in the Hamilton dynamics, so the issue in the Nambu dynamics under the influence of dissipation is open to us. In clarifying the behavior of the monodromy, the differential Galois theory is promising to give a group theoretical understanding of it as well as the emergence of chaos and/or fractal.
The detailed analysis within the framework of NNET remains an open problem, which will be addressed in future work\footnote{We intentionally leave the rigorous analysis of the necessary conditions for future work.}.

\color{black}
\paragraph{Applicability to high-dimensional and many-body systems}
\label{par:high-dim-many-body}

For higher-dimensional systems, explicitly carrying out the local reduction for all degrees of freedom
\( x \in \mathbb{R}^N \) and constructing the full set of Hamiltonians \( H_I \) and the entropy \( S \)
becomes computationally demanding in general.

From a practical viewpoint, it is therefore natural to apply the NNET formalism at the level of an effective dynamics for a small number of coarse-grained variables, such as order parameters, densities, fluxes, or energies, while treating the remaining degrees of freedom statistically. This strategy is particularly effective when a separation of time scales between slow and fast variables is present. In such cases, projection-operator techniques, including the Zwanzig and Mori--Zwanzig formalisms, allow the influence of fast variables to be incorporated into the reduced dynamics in the form of memory kernels and fluctuation terms.

In fact, in our previous work (Appendix C of \cite{paper 0}), we explicitly discussed a framework based on the Zwanzig projection method, where a slow/fast decomposition is performed and the contribution of fast variables is incorporated as stochastic fluctuations in an effective theory. The local reduction developed in the present work can be applied to such reduced equations involving a small number of effective degrees of freedom, and is therefore expected to be practically applicable to many-body systems.
\color{black}

\color{black}
Possible numerical validation strategies are briefly discussed\footnote{
For quasi-periodic systems, one may introduce a phase coordinate on \( S^{1} \) and reconstruct the entropy-like function \( S^{C} \) from observed velocity-field data by computing the divergence \( \nabla \cdot \dot{x} \) and solving
\( \Delta S^{C} = \nabla \cdot \dot{x} \).
Closed trajectories can then be fitted by circular or elliptical level sets to define \( H_{1} \), while the remaining Hamiltonians \( H_{I} \) are constructed locally via Darboux’s theorem.
The validity of the NNET reduction can be numerically assessed by checking:
(i) that the velocity field reconstructed from \( (H^{C}, S^{C}) \) reproduces the observed dynamics within a prescribed accuracy;
(ii) the consistency of the divergence relation \( \nabla \cdot \dot{x} \approx \Delta S^{C} \);
and (iii) the emergence of quasi-conserved quantities satisfying
\( \frac{d H_{I}^{C}}{dt} = \nabla H_{I}^{C} \cdot \nabla S^{C} \).
}.
\color{black}

\section*{Acknowledgments}
We would like to thank Toshio Fukumi about the non-linear response theory.
We are indebted to Shiro Komata and Ken Yokoyama for reading this paper and giving valuable comments. 

\appendix
\section*{Appendix A \color{black}: Physical view of Darboux theorem}

Darboux introduced two kinds of differentials (variations) for a $N$-dimensional vector field $X(x)_{k}$ ($k=1, \cdots, n$), $\Theta_d= \sum_{k=1}^n X_k(x) dx^k$ and $\Theta_{\delta}= \sum_{k=1}^N X_k(x) \delta x^k$;  the former differential $\Theta_d$ is induced by the temporal development of the coordinate $x_k=x_k(t)$ in an \emph{auxiliary} parameter $t$, while the latter differential $\Theta_{\delta}$ describes the variation (ensemble) of trajectories. So, the starting equation by Darboux, (10), $(10)^a, (10)^b$, and $(10)^c$ in \cite{Darboux}, is the following:
\begin{align}
\delta \Theta_d - d \Theta_{\delta}= \lambda    \Theta_{\delta} \; dt, \nonumber ~~\text{(Darboux)}~~~~~~~(A.1)
\end{align}
where $\lambda$ is a parameter, which can be anything, $\lambda=0, \ne 0$ or a function in $t$.

\subsubsection*{1) Darboux theory as electromagnetism}
If we consider the vector field $X_k(x)$ be a gauge potential $A_k(x)$ at a location $x_k(t)$ at time $t$ of a \emph{charged particle} with a unit of charge and mass, then (A.1) can be written as
\begin{align}
dx^i F_{ik}(\bm{x}(t)) \delta x^k = \lambda dt A_k(\bm{x}(t)) \delta x^k,  \nonumber ~~~~~~~~(A.2)
\end{align}
where the field strength is defined by $F_{\mu\nu}(x)=\partial_{\mu} A_{\nu}(x)-\partial_{\nu} A_{\mu}(x)$, where $(\mu\nu)=0, 1, 2, 3, (i,j,k)=1, 2, 3$.  $F_{\mu\nu}$ includes magnetic fields $\bm{B}$ and electric fields $\bm{E}$,
\begin{align}
F_{ij}=-\epsilon_{ijk_3\cdots k_N}\bm{B}^{k_3\cdots k_N}(=-\epsilon_{ijk}\bm{B}^k ~\text{for}~N=3), ~\text{and}~F_{0j}=-\bm{E}^j. \nonumber ~~(A.3)
\end{align}
Under a temporal gauge $A_0(x)=0$, we assume the following time dependency
\begin{align}
\bm{A}(\bm{x}, t)= e^{\lambda t} \bm{A}(\bm{x}), \nonumber  ~~~~~~~~~(A.4)
\end{align}
from which the electric field becomes $F_{0j}(x)=\partial_0 A_i(x)=\lambda A_i(x)= e^{\lambda t} \lambda A_i (\bm{x})$, and the magnetic field is $B_{ij}(x)= e^{\lambda t}  B_{ij} (\bm{x})$, so that (A.2) yields
\begin{align}
\frac{dx^i}{dt} F_{ik}(\bm{x}) \delta x^k =  F_{0k}(\bm{x}) \delta x^k.  \nonumber~~~~~~~~(A.5)
\end{align}
Comparison of (A.5) with the familiar formula of \emph{Lorentz force} tells us that the Darboux equation (A.1) in the Pfaff form is physically, the \emph{vanishing condition of Lorentz force} in electromagnetism.  

\subsubsection*{2) Invariance principle}
Darboux emphasized the \emph{invariance principle}, and utilized it for the general coordinate transformation (diffeomorphism) as well as the gauge transformation, in the terminology at present. The diffeomorphism is, of course, restricted to the coordinate transformation in $N$-spacial dimensions, without time.  So, the Darboux system is the non-relativistic electromagnetism. 

Viewing in this way, we can utilize, in addition to the Darboux equation, the \emph{equation of motion} of the charged particle. It reads in the original flat frame $x$,
\begin{align}
\ddot x^{i}(t) = F^i(\text{Lorentz})(x) = \delta^{ik} e^{\lambda t} \left( \frac{dx^j}{dt} F_{jk}(\bm{x}) +  F_{0k}(\bm{x}) \right)  =0. \nonumber ~~~~~~~~(A.6) 
\end{align}
If we move to another frame $y$, by coordinate transformation (diffeomorphism), $y^{i}=y^{i}(\bm{x})$, the new $y$-frame may be a curved space.  Then, with the metric $g'_{\mu\nu}(y)$, the invariance principle gives the equation of motion in the $y$ frame: 
\begin{align}
\ddot y^{i}(t)-\Gamma^i_{jk}\dot y^{j}\dot y^{k} = F^i(\text{Lorentz})(x) = g'^{ik}(y)e^{\lambda t} \left( \frac{dy^j}{dt} F_{jk}(\bm{y}) +  F_{0k}(\bm{y}) \right)  =0, \nonumber ~~(A.7) 
\end{align}
where $\Gamma^i_{jk}$ is the Christoffel's symbol.

The invariance principle connects two frames by
\begin{align}
&g'^{ij}(y)= \sum_a \frac{\partial y^{i}}{\partial x^{a}} \frac{\partial y^{j}}{\partial x^{a}} ~~(A.8a), ~~~ \dot{y}^i= \dot{x}^a \frac{\partial y^i}{\partial x^a}\nonumber ~~~~~(A.8b)\\
&F_{ab}(x)= F'_{ij}(y) \frac{\partial y^i}{\partial x^a} \frac{\partial y^j}{\partial x^b},  ~~
A_{a}(x)= A'_{i}(y) \frac{\partial y^i}{\partial x^a}. \nonumber ~~~(A.8c)
\end{align}
The solution of the equation of motion in the new frame gives the \emph{geodesic line}, satisfying 
\begin{align}
 \ddot y^{i}(t)-\Gamma^i_{jk}\dot y^{j}\dot y^{k}=0, ~ \nonumber ~~~~~~(A.9)
\end{align}
which corresponds to a straight line $\ddot{x}^{a}=0$ in the original frame.  In the electromagnetism of the Darboux system, there is no force acting on the charged particle, so that the trajectories form an ensemble of geodesic lines in both $x$- and $y$-frames.  During a short time ($t \ll 1$), we have
\begin{align}
&x^{a}(t)= x^{a} + v^{a}(x) t, ~\text{and also}~~y^{i}(t) = y^{i} + v'^{i}(y) t,  ~\nonumber ~~(A.10)
\end{align}
which gives $\dot{x}^{a}= v^{a}(\bm{x}), ~\dot{y}^{i} = v'^{i}(\bm{y})$, being approximately $t$-independent.

\subsubsection*{3) Darboux's solution}
For any field strength under the vanishing condition of Lorentz force, Darboux claims that we can find a new frame $y$ in which the field strength and gauge potential take simple forms, called \emph{canonical forms}.
The canonical forms (with the symbol $(C)$) found in \cite{Darboux} are written as follows, by fixing an integer $p$ ($2p \le N$):
\begin{align}
&F'^{(C)}_{ij}(\bm{y})= \sum_{k=1}^p \left( \delta_{i, 2k-1}\delta_{j, 2k}-\delta_{i, 2k}\delta_{j, 2k-1} \right): \; \text{constant}, \nonumber ~~~~(A.12) \\
&A'^{(C)}_{i}(\bm{y})=  \sum_{k=1}^p  \left( - \delta_{j, 2k-1}y^{2k} + \delta_{j, 2k} y^{2k-1} \right)F'^{(C)}_{ij}: \; \text{linear in}~y, \nonumber ~~~~(A.13) \\
&F'^{(C)}_{0i}(\bm{y}) = - \lambda A'^{(C)}_{i}(\bm{y}): \; \text{linear in}~y. \nonumber \hspace{4cm} (A.14)
\end{align}
He found these results by solving the differential equation (A.5) explicitly, which is  physically the vanishing condition of Lorentz force, 
\begin{align}
 \frac{dy^j}{dt} F_{ji}(\bm{y}) = \lambda A_i(\bm{y}).  \nonumber ~~~\text{(Darboux)}~~~~(A.15)
 \end{align}
Since we have an advantage to utilize the equation of motion, implying the trajectory be a geodesic line (A.10), the differential equation can be reduced to the easier equation in \emph{linear algebra},
 \begin{align}
  F'_{ij}(\bm{y}) v'^j(\bm{y})  =-\lambda A'_i(\bm{y}).  \nonumber ~~~~~~~(A.16)
 \end{align}
 Even in linear algebra, we have to distinguish two types det$F'_{ij}=0$ (indeterminant) and $\ne 0$ (determinant) as was done by Darboux, but in both cases, we can confirm his results.  Namely, $2p$ is the rank of the matrix $F'_{ij}$, ($i, j=1, \cdots, N$), and the algebraic solution can be written easily, 
 \begin{align}
 v'_i(\bm{y})= \begin{cases} 
 ~\lambda y^i~~ (i=1, \cdots, 2p), \\
 ~~0~~~~(i=2p+1, \cdots, N), 
 \end{cases}  \nonumber  ~~~~~(A.17)
 \end{align}
associated with the canonical forms, $F'_{ij}=F'^{(C)}_{ij}$ and $A'_i(\bm{y})= A'^{(C)}_{i}$.  Since $y^i$ is the initial position of the trajectory $y^i(t=0)=y^i$, the ensemble of trajectories satisfies $y^i(t)= (1+ \lambda t)y^i$ for $i \le 2p$ while $y^i(t)=y^i$ for $i \ge 2p$.  This ensemble seems physically allowable, only when $\lambda t \ll 1$ holds and the dynamics be \emph{static}, in which all the charged particles do not move $y^i(t)=y^i$ for $i=1, \dots, N$, and the magnetic field $\bm{B}$ dominates over the electric field $\bm{E}$.

Once we know the canonical form, it is easy to find the diffeomorphism to the new canonical frame $y^i=y^i (x)$, by using (A.8c) (A.13), 
\begin{align}
A_{a}(\bm{x})= A'^{(C)}_{i}(\bm{y}) \frac{\partial y^i}{\partial x^a}(\bm{x}), \nonumber ~~~~~~(A.18)
\end{align}
 and the metric of the new frame can be determined by (A.8a).
 
\subsubsection*{4) Examples for $N=3, 4$}
The result of Darboux theorem for $N=3$ is familiar in physics.  We know if we move to a new frame, the constant magnetic field can be expressed as $\bm{B}'(y)=(0, 0, b)$ with a constant $b$, and the corresponding gauge potential $\bm{A}'(y)$ is a 1-form field, $\bm{A}'_1= -\frac{1}{2}b (-y^2 dy^1 + y^1 dy^2) \to  y^{'1} dy^{'2} - y^{'2} dy^{'1}$, by a change of scale. The corresponding field strength is
\begin{align}
F'_{ij}(y)=  \begin{pmatrix} 0 & 1 &0 \\ -1 & 0 &0 \\
0 & 0 & 0 \end{pmatrix}. \nonumber ~~~~~(A.19)
\end{align}
This is an example of the Darboux frame for the indeterminant type with det$F_{ij}=0$.   

In the same manner, for $N=4$, we have $A'_{1}(y)=-y^2 dy^1 + y^1dy^2 -y^4 dy^3 + y^3d y^4$ and 
\begin{align}
F'_{ij}(y)=  \begin{pmatrix} 0 & 1 &0 & 0\\ -1 & 0 &0 &0 \\
0 & 0 & 0 & 1 \\
0 & 0& -1& 0 \end{pmatrix}. \nonumber ~~~~~(A.20)
\end{align}
This is the determinant type with det$F_{ij} \ne 0$.

\section*{Appendix B: Examples of the reduction}

In our papers including \cite{paper I}, \cite{paper III}, a number of examples show the reduction from a complex non-linear non-equilibrium systems to a simple Nambu non-equilibrium thermodynamics (NNET).

\subsubsection{1) Triangular chemical reaction}
This is a prototype of our formulation.  The general case of triangular reaction discussed in \cite{paper I}, which can be reduced to the following NNET with 
\begin{align}
&H^C_{1}=\frac{1}{4}(x_{1}^{2}+x_{2}^{2}+x_{3}^{2}), \\
&H^C_{2}=\sum_{(ijk) =1-3: \; \text{cyclic} }k_{ij}\left(1-\frac{k_{ji}}{k_{ij}}e^{\beta(A_{ij}-A_{(0)ij})}\right)x_{k},  \\
&S^C=-\frac{1}{2}\sum_{(ijk) =1-3: \;  \text{cyclic} } k_{ij}\left(1-\frac{k_{ji}}{k_{ij}}e^{\beta(A_{ij}-A_{(0)ij})}\right)\left(x_{i}^{2}-x_{i}x_{j}\right), 
\end{align}
where $x_i$ is the concentration of $i$-th atom, the affinity force $A_{ij}=  \mu_i - \mu_j$ is the difference of chemical potentials between different species of atoms, and $A_{(0)ij}$ is the equilibrium value at the new cycle $\Omega^C$.

For the triangular chemical reaction, an explicit construction of the Hamiltonians and the entropy by Helmholtz decomposition and Darboux theorem is presented in Appendix C.

\subsubsection{ 2) Belousov-Zhabotinski (BZ) reaction \cite{BZ} }
This is a typical chemical reaction of non-linear non-equilibrium system far from equilibrium, which gives a periodic color change of a bromine solution.  This is expressed in terms of $N=3$ NNET with three variables, $x=[$BrO$_{2}$], $y=[$Br$^{-}$], and $z=[$Ce$^{4+}$], and $(H_1, H_2, S)$.  A numerical analysis about the temporal change of $(H_1, H_2, S)$ as well a $(x, y, z)$ are studied in \cite{paper III}.  The oscillatory behavior of Hamiltonians and entropy can be seen having spikes.

This example illustrates how the existence proposition captures cyclic dynamics within the NNET framework.

\subsubsection{3) Hindmarsh-Rose (H-R), Lorenz 
and Chen models}
In the same manner in \cite{paper III}, Hindmarsh-Rose (H-R) model \cite{HR-1}\cite{HR-2} useful to describe the propagation of spiky signal along the nerve, and typical examples of generating chaos, Lorenz model \cite{LC-1} and Chen model \cite{LC-2}\cite{LC-3}, are also studied within the framework of NNET.   

The Hindmarsh-Rose model illustrates spike dynamics, while Lorenz and Chen provide chaotic examples; all can be reduced to NNET, demonstrating the broad applicability of the existence proposition.

These examples confirm that both oscillatory and chaotic behaviors can be embedded within the NNET framework, reinforcing the generality of the existence proposition.
Together, these examples demonstrate that NNET provides a unifying framework capable of reducing both oscillatory and chaotic systems to a simpler structure.

\color{black}
\section*{Appendix C Explicit Construction of $(H{{}_1},H{{}_2},S)$
\setcounter{section}{3}
\setcounter{equation}{0} 
in a Triangular Chemical Reaction within Nambu Non-equilibrium Thermodynamics}

This section provides a concrete example of applying the existence
theorem discussed in Nambu Non-equilibrium Thermodynamics to a triangular
chemical reaction. By employing the Helmholtz decomposition and Darboux
theorem, we explicitly construct the two Hamiltonians $H_{1}$, $H_{2}$
and the entropy $ $S that reproduce the dynamical equations of the
triangular reaction. This example demonstrates how the abstract existence
theorem can be realized in practice.

\section*{C.1 Nambu Non-equilibrium Thermodynamics and the Existence Theorem}

In Nambu Non-equilibrium Thermodynamics (NNET), a formal
existence theorem was introduced: any autonomous dynamical system
with both incompressible and compressible flows can, in principle,
be reduced to a NNET formulation with multiple Hamiltonians and an
entropy. The proof of this theorem relies on Helmholtz decomposition
and Darboux’s theorem, but it was presented only in a formal manner
without explicit calculations.

The purpose of this section is to provide a concrete realization
of this theorem. As a prototype example, we take the triangular chemical
reaction and explicitly construct the two Hamiltonians $H_{1}$, $H_{2}$
and the entropy $S$, showing that they reproduce the original dynamical
equations.

\section*{C.2 Formulation of Nambu Non-equilibrium Thermodynamics and the Existence
Theorem}

Nambu Non-equilibrium Thermodynamics (NNET) is formulated by extending
Nambu dynamics with multiple Hamiltonians and an entropy function.
For three thermodynamic variables $x_{i}(i=1,2,3)$, the equations
of motion take the form

\begin{equation}
\dot{x}^{i}=\{H_{1},H_{2},x^{i}\}+L_{ij}\frac{\partial S}{\partial x^{j}}
\end{equation}

where $\{A,B,C\}$ denotes the Nambu bracket defined by

\begin{equation}
\{A,B,C\}=\epsilon_{ijk}\frac{\partial A}{\partial x^{i}}\frac{\partial B}{\partial x^{j}}\frac{\partial C}{\partial x^{k}}
\end{equation}

Here, $H_{1}$ and $H_{2}$ represent Hamiltonians associated with
reversible dynamics, while $S$ is the entropy responsible for irreversible
dynamics. The transport coefficients $L_{ij}$ form a symmetric positive
semi-definite matrix.

In our discussion, a formal existence theorem was stated: given an
autonomous dynamical system consisting of incompressible and compressible
flows, one can, at least locally, construct Hamiltonians and an entropy
such that the system is represented in the above NNET form.

The proof of this theorem employs Helmholtz decomposition, which separates
any vector field into incompressible and compressible parts, and Darboux’s
theorem, which guarantees the local canonical form of two-forms. These
tools ensure that the reversible and irreversible contributions to
the dynamics can be consistently represented by Hamiltonians and entropy,
respectively.

\section*{C.3 Triangular Reaction as a Concrete Example}

As a concrete example, we now consider a triangular chemical reaction
and explicitly construct the functions $H_{1}$, $H_{2}$, $S$ after
applying the Helmholtz decomposition.

The triangular reaction is a cyclic chemical process, $X_{1}\to X_{2}\to X_{3}\to X_{1}$

\begin{equation}
\xymatrix@C=4em@R=4em{
X_{1} \ar@<0.5ex>[r]^{k_{12}} \ar@<0.5ex>[d]^{k_{13}} & X_{2} \ar@<0.5ex>[l]^{k_{21}} \ar@<0.5ex>[ld]^{k_{23}} \\
X_{3} \ar@<0.5ex>[u]^{k_{31}} \ar@<0.5ex>[ru]^{k_{32}} &
}
\end{equation}

which is described by the following set of rate equations:
\begin{equation}
\frac{dx_{1}}{dt}=-\tilde{k}_{12}x_{1}+\tilde{k}_{31}x_{3},
\end{equation}
\begin{equation}
\frac{dx_{2}}{dt}=-\tilde{k}_{23}x_{2}+\tilde{k}_{12}x_{1},
\end{equation}
\begin{equation}
\frac{dx_{3}}{dt}=-\tilde{k}_{31}x_{3}+\tilde{k}_{23}x_{2}.
\end{equation}

Here the coefficients, $\tilde{k}_{ij}$ are defined in terms of the
bare reaction rates $k_{ij}$ and the thermodynamic affinities $A_{i\to j}$
as
\begin{equation}
\tilde{k}_{ij}\equiv k_{ij}\left(1-\frac{k_{ji}}{k_{ij}}e^{\beta(A_{i\to j}-A_{(0)i\to j})}\right),
\end{equation}
where $\beta$ denotes the inverse temperature and $A_{i\to j}$ the
affinity between species $X_{i}$ and $X_{j}$, with $A_{i\to j}^{(0)}$
its equilibrium value.

We now apply the Helmholtz decomposition to separate the incompressible
and compressible parts of the flow, and construct the corresponding
Hamiltonians and entropy.

\section*{C.4 Helmholtz Decomposition}

In three dimensions, the Helmholtz decomposition allows any vector
field $V^{i}=\dot{x}^{i}$ to be expressed as
\begin{equation}
\dot{x}^{i}=\phi^{i}+B^{i}+\frac{\partial S}{\partial x^{i}}
\end{equation}
where $\phi^{i}$ satisfies $\Delta\phi^{i}=0$, $B^{i}$ is constructed
from an antisymmetric tensor $B_{ij}$ as $B^{i}=\epsilon^{ijk}B_{jk}$,
and $\frac{\partial S}{\partial x^{i}}$ is derived from a scalar
potential $S$.

Let us determine $\phi$, $B^{i}$, and $S$ that satisfy this condition
from the triangular reaction. For simplicity, we first set $\phi=0$.

Next, we determine $S$ and then obtain $B^{i}$ from $V^{i}-\frac{\partial S}{\partial x^{i}}$.

First, let us note that

\begin{equation}
\nabla\cdot V=\Delta S=-k_{\Sigma},
\end{equation}

with
\begin{equation}
k_{\Sigma}\equiv\tilde{k}_{12}+\tilde{k}_{23}+\tilde{k}_{31}.
\end{equation}

Next, we observe that the triangular reaction has a conserved quantity
\begin{equation}
x_{1}+x_{2}+x_{3},
\end{equation}

and we set
\begin{equation}
H_{1}\equiv x_{1}+x_{2}+x_{3}.
\end{equation}

Since $H_{1}$ is a conserved quantity, it must satisfy
\begin{equation}
\nabla H_{1}\cdot V=0.
\end{equation}

A function $S$ that fulfills this condition can be chosen as
\begin{equation}
\nabla S=\frac{k_{\Sigma}}{2}\left(-x_{1}+x_{2},-x_{2}+x_{1},0\right),
\end{equation}

which gives

\begin{equation}
S=-\frac{k_{\Sigma}}{4}(x_{1}-x_{2})^{2}.
\end{equation}

From this, we can determine $B^{i}$ as

\begin{equation}
B^{i}=\left(\begin{array}{ccc}
-\left(\tilde{k}_{12}-\frac{k_{\Sigma}}{2}\right) & -\frac{k_{\Sigma}}{2} & \tilde{k}_{31}\\
\left(\tilde{k}_{12}-\frac{k_{\Sigma}}{2}\right) & -\left(\tilde{k}_{23}-\frac{k_{\Sigma}}{2}\right) & 0\\
0 & +\tilde{k}_{23} & -\tilde{k}_{31}
\end{array}\right)\left(\begin{array}{c}
x_{1}\\
x_{2}\\
x_{3}
\end{array}\right).
\end{equation}

\section*{C.5 Darboux Part}

Next, assuming that Darboux’s theorem holds, we determine $H_{2}$.

We have
\begin{equation}
B=\nabla H_{1}\times\nabla H_{2}=\left(\begin{array}{c}
1\\
1\\
1
\end{array}\right)\times\nabla H_{2}.
\end{equation}

Introducing a matrix $M$, we require

\begin{equation}
\left(\begin{array}{c}
1\\
1\\
1
\end{array}\right)\times\nabla H_{2}=M\left(\begin{array}{c}
x_{1}\\
x_{2}\\
x_{3}
\end{array}\right),
\end{equation}
which yields
\begin{equation}
M=\left(\begin{array}{ccc}
a & b & c\\
a & b+\tilde{k}_{23} & c-\tilde{k}_{31}\\
a-\tilde{k}_{12}+\frac{k_{\Sigma}}{2} & b+\tilde{k}_{23}-\frac{k_{\Sigma}}{2} & c
\end{array}\right).
\end{equation}

Here, $a,b,c$ are not independent but constrained by
\begin{equation}
a=b=c+\frac{\tilde{k}_{12}-\tilde{k}_{23}-\tilde{k}_{31}}{2}
\end{equation}

By choosing
\begin{equation}
c=-\frac{\tilde{k}_{12}-\tilde{k}_{23}-\tilde{k}_{31}}{2}
\end{equation}

so that $a,b$ vanish, we obtain
\begin{equation}
H_{2}=\frac{1}{2}\left(c-\tilde{k}_{12}+\frac{k_{\Sigma}}{2}\right)x_{3}x_{1}+\frac{1}{2}\tilde{k}_{23}x_{2}^{2}+\frac{1}{2}\left(c-\tilde{k}_{31}+\tilde{k}_{23}-\frac{k_{\Sigma}}{2}\right)x_{2}x_{3}.
\end{equation}

Thus, for the triangular reaction we obtain the set($H_{1},H_{2},S$)
\begin{equation}
H_{1}\equiv x_{1}+x_{2}+x_{3},
\end{equation}

\begin{equation}
H_{2}=\frac{1}{2}\left(c-\tilde{k}_{12}+\frac{k_{\Sigma}}{2}\right)x_{3}x_{1}+\frac{1}{2}\tilde{k}_{23}x_{2}^{2}+\frac{1}{2}\left(c-\tilde{k}_{31}+\tilde{k}_{23}-\frac{k_{\Sigma}}{2}\right)x_{2}x_{3},
\end{equation}

\begin{equation}
S=-\frac{k_{\Sigma}}{4}\left(x_{1}-x_{2}\right)^{2},
\end{equation}

with
\begin{equation}
c=-\frac{\tilde{k}_{12}-\tilde{k}_{23}-\tilde{k}_{31}}{2}
\end{equation}

Remark:

It is also possible not to take $H_{1}$ as a conserved quantity. For
example, one may choose

\begin{equation}
H_{1}=x_{1}^{2}+x_{2}^{2}+x_{3}^{2}
\end{equation}

in which case $H_{2}$ and $S$ take symmetric forms. Which choice
is appropriate depends on the situation.
\color{black}

\color{black}
\section{Time Evolution of Entropy Production in Nambu Non-equilibrium Thermodynamics}
\label{app:entropy-stability}

In this appendix, we present a linear stability analysis within the framework of Nambu Non-equilibrium Thermodynamics (NNET).
We discuss how stability criteria are expressed in terms of the Hamiltonians \( H_{1},\dots,H_{N-1} \) and the entropy-like function \( S \),
and derive conditions for structural stability arising from their interplay.

\subsection{Linear Stability Analysis}
\label{app:linear-stability}

Consider an autonomous dynamical system
\begin{equation}
\frac{d x_{i}}{dt} = v_{i}(\bm{x}) .
\end{equation}
Let \( \bm{x}_{0} \) be a stationary solution.
A small perturbation \( \delta \bm{x} \) is introduced as
\begin{equation}
x_{i} = x_{0,i} + \delta x_{i}.
\end{equation}
Substituting into the equation of motion and linearizing, we obtain
\begin{equation}
\frac{d \delta x^{i}}{dt}
= \frac{\partial v_{i}}{\partial x^{j}}(\bm{x}_{0})\, \delta x^{j}
\equiv \Lambda_{ij}\, \delta x^{j}.
\end{equation}

The general solution is expressed in terms of the eigenvalues \( \lambda_{a} \) and eigenvectors \( \psi_{a}^{i} \) of \( \Lambda \):
\begin{equation}
\delta x^{i}(t) = \sum_{a} c_{a} e^{\lambda_{a} t} \psi_{a}^{i},
\end{equation}
where the coefficients \( c_{a} \) are determined by the initial perturbation.
If at least one eigenvalue has a positive real part, the stationary solution is linearly unstable.

For a two-dimensional system, the eigenvalues are explicitly given by
\begin{equation}
\lambda
= \frac{1}{2}
\left(
\Lambda_{11} + \Lambda_{22}
\pm
\sqrt{(\Lambda_{11}-\Lambda_{22})^{2}
+ 4 \Lambda_{12}\Lambda_{21}}
\right).
\end{equation}

\subsection{Onsager Theory}
\label{app:onsager}

In Onsager’s theory, the dynamics is governed by
\begin{equation}
\dot{x}^{i} = L_{ij} \frac{\partial S}{\partial x^{j}},
\qquad
L_{ij} = L_{ji}.
\end{equation}
The entropy production rate is
\begin{equation}
\dot{S}
= L_{ij} \frac{\partial S}{\partial x^{i}} \frac{\partial S}{\partial x^{j}}
\ge 0,
\end{equation}
which implies that \( L_{ij} \) is positive semi-definite.

Near equilibrium, stability requires
\begin{equation}
\ddot{S}
= 2
\frac{\partial^{2} S}{\partial x^{i} \partial x^{k}}
L_{ij} \frac{\partial S}{\partial x^{j}}
L_{kl} \frac{\partial S}{\partial x^{l}}
\le 0,
\end{equation}
so that the Hessian of \( S \) is negative semi-definite.
The linearized dynamics reads
\begin{equation}
\delta \dot{x}^{i}
= L_{ij} \frac{\partial^{2} S}{\partial x^{j} \partial x^{k}}
\delta x^{k},
\end{equation}
leading to
\begin{equation}
\Lambda_{ik}
= L_{ij} \frac{\partial^{2} S}{\partial x^{j} \partial x^{k}}.
\end{equation}

If \( L_{ij} \) and \( \partial_{i}\partial_{j} S \) commute,
\( \Lambda \) becomes negative semi-definite and the system is linearly stable.

\subsection{Hamiltonian Dynamics}
\label{app:hamiltonian}

For Hamiltonian dynamics,
\begin{equation}
\dot{x}^{i}
= \{x^{i},H\}
= \epsilon^{ij} \frac{\partial H}{\partial x^{j}} .
\end{equation}
Linearization yields
\begin{equation}
\Lambda_{ik}
= \epsilon^{ij} \frac{\partial^{2} H}{\partial x^{j} \partial x^{k}} .
\end{equation}

For a Hamiltonian of the form
\( H = \frac{p^{2}}{2m} + V(x) \),
the eigenvalues are
\begin{equation}
\lambda = \pm \sqrt{-\frac{1}{m} \frac{\partial^{2} V}{\partial x^{2}}}.
\end{equation}
Thus, harmonic potentials lead to purely imaginary eigenvalues and stable oscillations,
whereas inverted potentials yield unstable modes.

\subsection{Nambu Dynamics}
\label{app:nambu}

For three variables, Nambu dynamics is given by
\begin{equation}
\dot{x}^{i}
= \{x^{i}, H_{1}, H_{2}\}
= \epsilon^{ijk}
\frac{\partial H_{1}}{\partial x^{j}}
\frac{\partial H_{2}}{\partial x^{k}} .
\end{equation}
Linearization leads to
\begin{equation}
\Lambda_{il}
= \epsilon^{ijk}
\left(
\frac{\partial^{2} H_{1}}{\partial x^{j} \partial x^{l}}
\frac{\partial H_{2}}{\partial x^{k}}
+
\frac{\partial H_{1}}{\partial x^{j}}
\frac{\partial^{2} H_{2}}{\partial x^{k} \partial x^{l}}
\right).
\end{equation}

For quadratic Hamiltonians \( H_{1} = \sum_{i} \tfrac{1}{2} k_{i} x_{i}^{2} \)
and linear \( H_{2} = a_{i} x^{i} \),
the eigenvalues are
\begin{equation}
\lambda
= 0,\quad
\pm \sqrt{
- a_{3}^{2} k_{1} k_{2}
- a_{2}^{2} k_{1} k_{3}
- a_{1}^{2} k_{2} k_{3}
}.
\end{equation}

\subsection{Nambu Non-equilibrium Thermodynamics}
\label{app:nnet}

In NNET, the dynamics reads
\begin{equation}
\dot{x}^{i}
= \{x^{i}, H_{1}, H_{2}\}
+ \frac{\partial S}{\partial x^{i}} .
\end{equation}
The linearized operator becomes
\begin{equation}
\Lambda_{il}
= \epsilon^{ijk}
\left(
\frac{\partial^{2} H_{1}}{\partial x^{j} \partial x^{l}}
\frac{\partial H_{2}}{\partial x^{k}}
+
\frac{\partial H_{1}}{\partial x^{j}}
\frac{\partial^{2} H_{2}}{\partial x^{k} \partial x^{l}}
\right)
+ \frac{\partial^{2} S}{\partial x^{i} \partial x^{l}} .
\end{equation}

This shows that Hamiltonian contributions can either stabilize or destabilize entropy-driven dynamics,
in contrast to purely Onsager systems.

\subsection{Time Evolution of Entropy}
\label{app:entropy-evolution}

The entropy production rate is
\begin{equation}
\dot{S}
= \nabla S \cdot \nabla S
+ \nabla S \cdot (\nabla H_{1} \times \nabla H_{2}),
\end{equation}
and its time derivative reads
\begin{align}
\ddot{S}
=&\;
2 \frac{\partial^{2} S}{\partial x^{i} \partial x^{j}}
\frac{\partial S}{\partial x^{i}}
\frac{\partial S}{\partial x^{j}}
+
2 \frac{\partial^{2} S}{\partial x^{i} \partial x^{j}}
\frac{\partial S}{\partial x^{i}}
(\nabla H_{1} \times \nabla H_{2})^{j}
\nonumber \\
&+
\frac{\partial S}{\partial x^{i}}
\frac{\partial S}{\partial x^{j}}
\partial_{i} (\nabla H_{1} \times \nabla H_{2})^{j}
+
(\nabla H_{1} \times \nabla H_{2})^{i}
\partial_{i} (\nabla H_{1} \times \nabla H_{2})^{j}
\frac{\partial S}{\partial x^{j}} .
\end{align}

A sufficient condition for stability is that this quantity remains non-positive.
\color{black}

\end{document}